\documentclass[12pt]{article}
\usepackage[dvips]{graphicx}

\begin{document}


\begin{center}

{\LARGE Plasmon transport in graphene investigated by time-resolved measurement}

\vspace*{12pt}

{\large N. Kumada$^{1, \ast}$, S. Tanabe$^{1}$, H. Hibino$^{1}$, H. Kamata$^{1,2}$, M. Hashisaka$^{2}$, K. Muraki$^{1}$ and T. Fujisawa$^{2}$}
\vspace*{6pt}

\end{center}

$^1$NTT Basic Research Laboratories, NTT Corporation, 3-1 Morinosato-Wakamiya, Atsugi, Kanagawa, Japan\\
$^2$Department of Physics, Tokyo Institute of Technology, 2-12-1 Ookayama, Meguro, Tokyo, Japan


\newpage

\baselineskip10pt\small

{\bf
Plasmons, which are collective charge oscillations, offer the potential to use optical signals in nano-scale electric circuits \cite{Barnes,Maier,Ozbay}.
Recently, plasmonics using graphene \cite{Ryzhii,Rana,Jablan2009,Mishchenko,Ju} have attracted interest, particularly because of the tunable plasmon frequency through the carrier density $n$ \cite{Ju,Hwang2007}.
However, the $n$ dependence of the plasmon velocity is weak ($\propto n^{1/4}$) and it is difficult to tune the frequency over orders of magnitude.
Here, we demonstrate that the velocity of plasmons in graphene can be changed over two orders of magnitude by applying a magnetic field $B$ and by the presence/absence of a gate; at high $B$, edge magnetoplasmons (EMPs), which are plasmons localized at the sample edge, are formed and their velocity depends on $B$ and the gate screening effect.
The wide range tunability of the velocity and the observed low-loss plasmon transport encourage designing graphene nanostructures for plasmonics applications.
}

On metal surface, resonant interactions between electrons in nanoparticles and the electromagnetic field of light create surface plasmons.
Since the wavelength of surface plasmons can be reduced to the size dimensions of electronic devices, plasmonics have been proposed to merge optics and electronics \cite{Barnes,Maier,Ozbay}.
However, there are fundamental obstacles to use metals for plasmonics: plasmon properties cannot be tuned once the material and the device geometry are determined and the plasmon decay time is short ($\sim 10$\ fs).
In graphene, theory has predicted that the dispersion of sheet plasmons changes with the carrier density as $n^{1/4}$ \cite{Hwang2007}.
Recently, the tunable resonant frequency of sheet plasmons in graphene micro-ribbon structures have been demonstrated \cite{Ju}.
On the analogue of conventional two-dimensional systems in GaAs, edge magnetoplasmons (EMPs) are expected to be formed at high $B$ and the dispersion is also changed by applying $B$ \cite{Volkov,Aleiner,Ashoori,Zhitenev} and by the presence of a gate \cite{KumadaEMP}.
Furthermore, low-loss plasmon transport is expected \cite{Jablan2009} when the plasmon-electron \cite{Bostwick2007,Bostwick2010,Polini} and plasmon-phonon \cite{YuLiu2010,Hwang2010,Jablan2011} couplings are suppressed.
These studies suggest that graphene is a promising candidate material for the plasmonics.

Here, we show that the velocity of plasmons can be tuned over a wide range between 2,500 and 10\ km/s by $B$ and the gate using time-resolved transport measurements.
In the absence of $B$ and a gate, the velocity is faster than the Fermi velocity $v_{\rm F} \sim 1,000$\ km/s, evidence for the transport of sheet plasmons.
In quantum Hall (QH) effect regime at high $B$, the velocity decreases with the Hall conductance $\sigma _{xy}$, indicating the formation of EMPs, which are plasmons confined in QH edge channels.
In a sample with a gate, the screening effect reduces the velocity by one order of magnitude and changes the $B$ dependence of the EMP velocity.
Since edge channels formed at high $B$ and p-n junctions \cite{Mishchenko,Williams} formed by using gates are useful to guide plasmons, the wide-range tunability of the plasmon velocity using $B$ and a gate encourage designing graphene nanostructures for plasmonic circuits.

For this work, large-area graphene is essential to obtain long time of flight that can be resolved by electrical means.
We prepared a graphene wafer by thermal decomposition of a 4H-SiC(0001) substrate \cite{Tanabe}.
The edge of graphene devices was defined by etching the graphene and the SiC substrate.
The surface of the devices was covered with 100-nm-thick hydrogen silsequioxane (HSQ) and 40-nm-thick SiO$_2$ insulating layers.
As a result of n-doping from the SiC substrate and p-doping from the HSQ layer, graphene has n-type carriers with $n=5\times 10^{11}$\ cm$^{-2}$.
Longitudinal $R_{xx}$ and Hall $R_{xy}$ resistances in a millimeter-scale Hall bar show well developed $\nu =2$ and 6 QH states (Fig.\ 1{\bf b}), demonstrating that the carrier density is almost uniform even in the large device.

For the time-resolved transport measurement, we used two samples, one with and the other without a large top gate (Fig.\ 1{\bf c}).
Excited charges with the energy of $<1$\ meV are injected into graphene by applying a square voltage pulse to the injection gate deposited across the sample edge: at the rising and falling edges of the pulse, positive and negative charges are generated, respectively (Fig.\ 1{\bf d}).
The charges propagate in the sample and are detected as the time-dependent current by a sampling oscilloscope through the detector Ohmic contact, which is located 1.1\ mm away from the injection gate.
The origin of the time is set at the onset of the injection pulse (supplementary information).
All measurements were performed at 1.5\ K.

Figure\ 2 shows results for the sample without a top gate.
The current traces are asymmetric with respect to $B=0$\ T (Fig.\ 2{\bf a}) because of the chirality of the edge current: when $B$ is applied from the back of the sample ($B>0$), the chirality is clockwise and the injected charges travel to the detector Ohmic contact along the left edge (Fig.\ 1{\bf c}); otherwise ($B<0$), the charges flow to other grounded Ohmic contacts.
For $B>0$, detailed measurements with fine time and $B$ steps (Fig.\ 2{\bf b}) reveal that the amplitude and the time delay of the current pulse depend on $B$.
The amplitude of the current pulse is large around $B=3$\ T and for $B>5$\ T, where the $\nu =6$ and $\nu =2$ QH states are formed, respectively.
This indicates that the charge relaxation is mainly due to scattering by electrons in bulk graphene.
The time delay at the current peak, which corresponds to the time of flight of charges, is about 1\ ns.
It increases with $B$ and becomes almost constant in the $\nu =2$ QH state for $B>5$\ T \cite{widenu2}.

The velocity of charges can be calculated from the time of flight and the length of the edge (Fig.\ 2{\bf c}).
Note that for smaller $B$, where edge channels are not well developed, excited charges propagate in bulk graphene from the injection gate to all the Ohmic contacts.
The detected signal corresponds to charges drawn to the detector Ohmic contact.
Since we used the direct path length between the injection gate and the detector Ohmic contact to calculate the velocity, the value for smaller $B$ would be underestimated.
Around $B=0$\ T, the velocity is about 2,000\ km/s or larger, which is larger than $v_{\rm F}\sim 1,000$\ km/s.
This demonstrates that charges propagate not as individual electrons but as collective modes, that is, sheet plasmons.
Indeed, the velocity of sheet plasmons is calculated to be inversely proportional to $\sqrt {k}$ with $k$ the wave number \cite{Hwang2007}, and it is larger than $v_{\rm F}$ at small $k\sim 10^3$\ m$^{-1}$, which is relevant to the transport measurement.
As $B$ is increased, the velocity decreases with plateau structures appearing for the $\nu =6$ and $\nu =2$ QH states, suggesting that the velocity is a function of $\sigma _{xy}$.
This feature is an indicator of EMPs \cite{Volkov,Aleiner,Ashoori}; at high $B$, sheet plasmons have a gap corresponding to the cyclotron energy and gapless plasmons exist only in edge channels.

For a quantitative analysis, the velocity is compared with theory and experiment for GaAs QH systems.
In a GaAs QH system without a top gate, the velocity of EMPs is about 1,700\ km/s at $\nu =2$ and 5,000\ km/s at $\nu =6$ (inset of Fig.\ 2{\bf c}).
Theoretically, the velocity is given by \cite{Aleiner}
\begin{equation}
v=[{\rm ln}(e^{-\gamma }/2kw)-1]\sigma _{xy}/\epsilon,
\label{equngate}
\end{equation}
where $\gamma $ is the Euler constant and $\epsilon $ is the dielectric constant ($\epsilon _{\rm GaAs}=12.9 \epsilon _0$ in GaAs).
In this model, $w$ represents the transverse width of the edge potential and EMPs are confined within $w$.
Equation\ (\ref{equngate}) well fits the experimental result with a constant $w\sim 2$\ $\mu $m, which is consistent with a soft wall edge potential in GaAs.
In graphene, naively, the edge potential is a hard wall, where $w$ in equation\ (\ref{equngate}) is replaced by a length $l=e^2\nu /\epsilon \hbar \omega _c$ determined by the Coulomb energy and the cyclotron energy $\hbar \omega _c$ \cite{Volkov}: in our sample parameters, $l\sim 100$\ nm, which is much smaller than $w\sim 2$\ $\mu $m in GaAs.
Since $v$ in equation\ (\ref{equngate}) increases with decreasing $w$, EMPs are expected to be much faster than those in GaAs.
However, the velocities of 2,000\ km/s at $\nu =6$ and of 1,000\ km/s at $\nu =2$ are about half as large as those in GaAs.
If we use the dielectric constant of graphene $\epsilon =(\epsilon _{\rm SiC}+\epsilon _{\rm HSQ})/2=6.2\epsilon _0$, which is the average of the values of the HSQ insulating layer ($\epsilon _{\rm HSQ}=2.8 \epsilon _0$) and the SiC substrate ($\epsilon _{\rm SiC}=9.6 \epsilon _0$), the best fit is obtained by adjusting $w=12$\ $\mu $m.
This suggests that edge potential in our graphene devices is rather soft and/or EMPs are slowed down by some mechanisms that are not incorporated in equation\ (\ref{equngate}).

We estimate $w$ in our devices.
At the sample edge, since SiC is mesa-etched and the side of the mesa is covered with HSQ (Fig.\ 2{\bf d}), p-doping from HQS is predominant over the n-doping from SiC.
As a result, the potential for electrons increases gradually near the sample edge and $w$ would be larger than $l$.
On the other hand, the upper limit of $w$ is set by a DC transport measurement in a small Hall bar made by the same fabrication process \cite{Tanabe}: in a Hall bar with the width of 2.5\ $\mu $m, $R_{xx}$ and $R_{xy}$ show well developed $\nu =2$ and 6 QH states with $n=6\times 10^{11}$\ cm$^{-2}$, similar to those in a large Hall bar (Fig.\ 1{\bf b}).
This indicates that $w$ must be smaller than 1\ $\mu $m.
The maximum $w=1$\ $\mu $m is still much smaller than $w=12$\ $\mu $m obtained by the fitting.
This discrepancy demonstrates the existence of mechanisms that slow down EMPs.
One possible mechanism is the screening by charges in dopants.
Since dopants exist very close to graphene at a distance of $\sim 0.3$\ nm, a small change in positions of charges can partially screen the electric field of plasmons.
This effectively enlarges the dielectric constant and reduces the velocity of plasmons.
Note that although plasmon-phonon \cite{YuLiu2010,Hwang2010,Jablan2011} and plasmon-electron \cite{Bostwick2007,Bostwick2010,Polini} couplings also modify the velocity of plasmons, the couplings must be small for low-energy plasmons and they cannot be the only cause of the discrepancy.

Figure\ 3 shows results for the sample with a top gate, which demonstrate that the plasmon velocity can be further changed by the gate.
The behavior of the current pulse for the top gate bias $V_{\rm tg}=0$\ V (Fig.\ 3{\bf a}) is largely different from that for the ungated sample.
Typical time of flight is 10\ ns, which is one order of magnitude larger than that in the ungated sample.
In the $\nu =2$ QH state for $B>6$\ T, the time of flight increases with decreasing $B$.
Around $B=3$\ T, weak signal for the $\nu =6$ QH state appears.
The amplitude of the current pulse is small for $B$ away from QH states.
Similar measurements for $30\geq V_{\rm tg}\geq -10$\ V were carried out, and the velocity is plotted in Figs.\ 3{\bf b}-{\bf f}.
For $V_{\rm tg}=30$\ V, the velocity oscillates around 100\ km/s with peaks at $B=10.8$ and 6.2\ T, where the $\nu =6$ and 10 QH states are formed, respectively.
As $V_{\rm tg}$ is decreased, the field position of the peaks shifts to lower $B$ following the $\nu $ and, at the same time, the peak velocity decreases (Fig.\ 3{\bf g}).

The smaller velocity is due to the screening of the electric field in plasmons by the gate.
The degree of the screening is evaluated by $d/w$ with the gate-graphene distance $d$ and the velocity is given by \cite{Zhitenev1995,Johnson}
\begin{equation}
v=\sigma _{xy}d/\epsilon w.
\label{eqgate}
\end{equation}
Note that this model is developed for a system without dissipation and valid only around QH states in our device.
If we calculate $w$ using $d/\epsilon =d _{\rm HSQ}/\epsilon _{\rm HSQ}+d _{\rm SiO_2}/\epsilon _{\rm SiO_2}$ with $\epsilon _{\rm SiO_2}=3.9\epsilon _0$, $d _{\rm HSQ}=100$\ nm and $d _{\rm SiO_2}=40$\ nm, typical $w$ at $\nu =6$ becomes $\sim 10$\ $\mu $m.
The unrealistically large $w$ is similar to the result for the ungated sample (Fig.\ 2{\bf d}), again suggesting that the velocity of plasmons is reduced by interactions with their environment.

Meanwhile, the observed $\nu $ and $n$ dependence of the velocity cannot be explained by the $B$ dependence of $\sigma _{xy}$; rather, the oscillating behavior suggests a contribution of $\sigma _{xx}$.
Theory with the dissipation taken into account suggests that, in a gated two-dimensional electron system, the dissipation damps EMPs and, at the same time, slows EMPs down \cite{Johnson}.
This is consistent with the oscillation of the velocity as a function of $\nu $ with peaks in QH states.
The increase in the velocity at $\nu =6$ with $n$ suggests that the velocity increases with the gap of the QH state.

We demonstrated that plasmons propagate a distance of 1.1\ mm with the velocity depending on $B$, $n$ and the presence or absence of a gate.
Quantitative analyses of the velocity suggested that interactions with dopants and dissipations slow plasmons down.
This suggests that control of the environment of plasmons further increase the tunable range of the velocity.
The wide-range tunability of the velocity and the observed large time of flight indicate that graphene is a promising material for plasmonics applications.
Information on effects of $B$ and a gate is useful to design plasmonics devices.

{\bf Methods}

We prepared a graphene wafer by thermal decomposition of a 4H-SiC(0001) substrate.
SiC substrates were annealed at around 1,800\ $^{\circ }$C in Ar at a pressure of less than 100\ Torr \cite{Tanabe}.
For the fabrication of devices, graphene and the SiC were mesa etched in a CF$_4$/O$_2$ atmosphere.
After the etching, Cr/Au electrodes were deposited and then the surface was covered with 100-nm-thick HSQ and 40-nm-thick SiO$_2$ insulating layers.
For the injection gate and the top gate, Cr/Au was deposited on the insulating layers.
Although steps of the SiC substrate have been reported not to affect the plasmon dispersion \cite{Langer}, to minimize this possible effect, the edge between the injection gate and the detector Ohmic contact is aligned parallel to the substrate steps.

{\bf Acknowledgement}

The authors are grateful to K. Takase and K. Sasaki for fruitful discussions and to M. Ueki for experimental support.
This work was supported in part by Grant-in-Aid for Scientific Research (21000004) and (21246006) from MEXT of Japan.

{\bf Author Contributions}

N. K. performed experiments, analyzed data and wrote the manuscript.
S. T. and H. H. grew the wafer.
N. K, H. K., M. H., K. M., and T. F. discussed the results.
All authors commented on the manuscript.

\newpage

\begin{figure}[t]
\begin{center}
\includegraphics[width=0.5\linewidth]{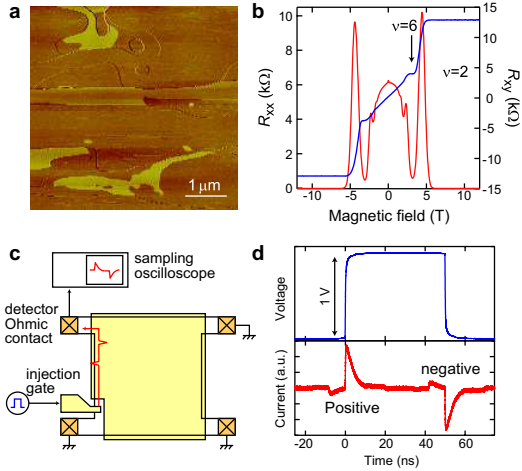}
\caption{{\bf Graphene on SiC and the experimental techniques.}
{\bf a}, Atomic force microscopy phase image of graphene on SiC.
Single-layer graphene (dark region) covers the substrate, while two or more graphene layers (bright regions) are formed along the terrace edge.
Since few-layer graphene regions are fragmented, the single-layer graphene dominates carrier transport.
{\bf b}, $R_{xx}$ and $R_{xy}$ at 1.5\ K of a Hall bar device with the channel width and length of 0.2 and 1.1\ mm, respectively.
The mobility is 12,000\ cm$^2$/Vs.
{\bf c}, Schematic illustration (not to scale) of the sample structure and the experimental setup for the time-resolved transport measurement.
In the ungated sample, the length of the edge is 1.1\ mm.
In the gated sample, the lengths in the regions with and without the top gate are 0.8 and 0.3\ mm, respectively.
{\bf d}, Voltage pulse with the amplitude of 1\ V and the time width of 50\ ns applied to the injection gate (top) and temporal traces of the current for $B=0$\ T detected by a sampling oscilloscope (bottom).
The current is averaged over a few seconds on the period of the injection pulse.
Small features around $-10$ and 40\ ns are due to reflections in high-frequency lines.}
\label{AFM}
\end{center}
\end{figure}

\clearpage

\begin{figure*}[t]
\begin{center}
\includegraphics[width=1.0\linewidth]{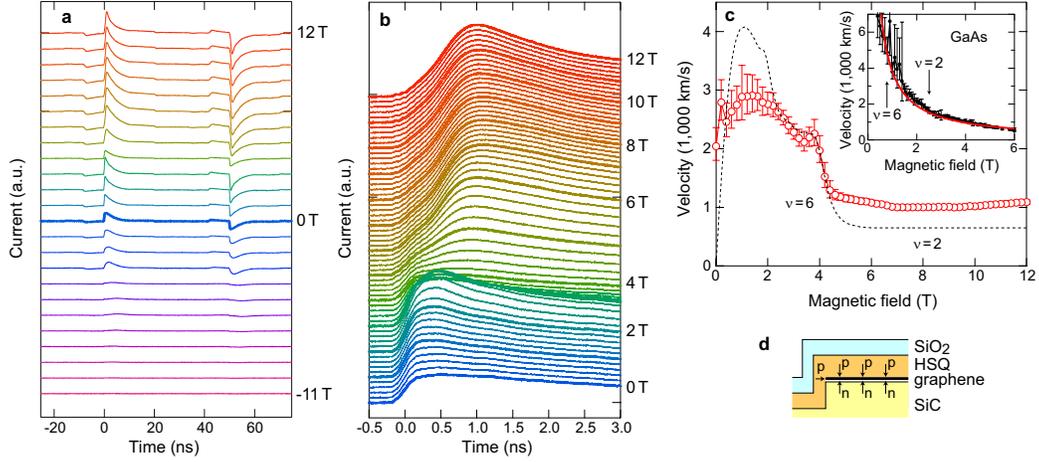}
\caption{{\bf Results of the time-resolved transport measurement for the graphene device without a top gate.}
{\bf a}, Current as a function of time for several magnetic fields between $B=-11$ (bottom) and 12\ T (top).
The sign of $B$ is defined as positive when $B$ is applied from the back of the sample.
The time width and repetition time of the injection pulse are 50 and 100\ ns, respectively.
Traces are vertically offset for clarity.
{\bf b}, Current as a function of time for positive magnetic fields with fine steps and the time range around the positive charge transport.
{\bf c}, Velocity of charges calculated from the measured time of flight and the length of the edge (1.1\ mm).
The dashed line represents the velocity calculated using equation\ (\ref{equngate}) with $w=12$\ $\mu $m; for the calculation, we used $\sigma _{xy}$ measured in the Hall bar sample and $k=2\pi/1.1$\ mm$^{-1}$ with 1.1\ mm the length of the edge.
The inset shows the velocity of EMPs in a GaAs QH system taken from Ref.\ \cite{KumadaEMP}.
The red line is the result of fitting with $w\sim 2$\ $\mu $m.
{\bf d}, Schematic illustration of the non-uniform doping near the sample edge.}
\label{ungate}
\end{center}
\end{figure*}

\clearpage

\begin{figure*}[t]
\begin{center}
\includegraphics[width=0.95\linewidth]{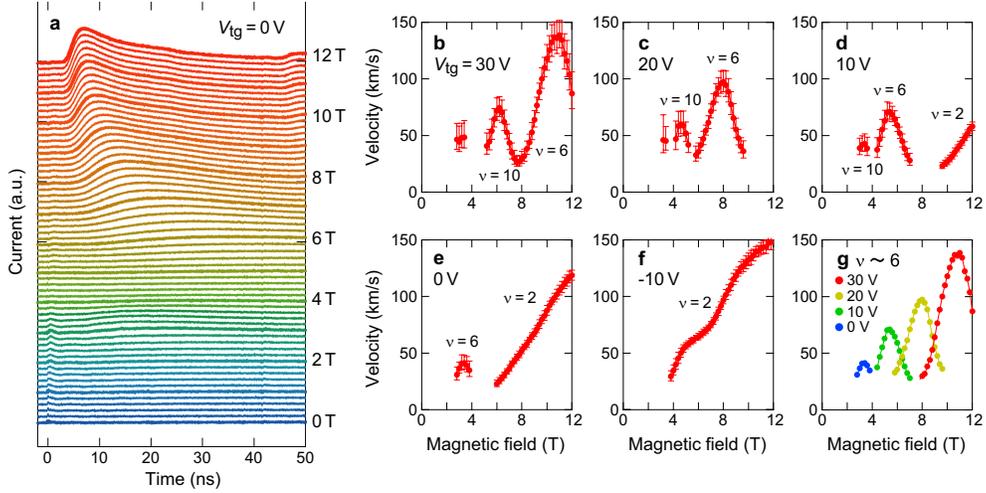}
\caption{{\bf Results of the time-resolved transport measurement for the graphene device with a top gate.}
{\bf a}, Current as a function of time for $V_{\rm tg}=0$\ V for positive magnetic fields and the time range around the positive charge transport.
The time width and repetition time of the injection pulse are 100 and 200\ ns, respectively.
Traces are vertically offset for clarity.
{\bf b}-{\bf f}, The velocity of charges in the gated region for several values of the top gate bias between 30 and $-10$\ V.
The charge neutrality point is located at $V_{\rm tg}=-40$\ V and thus the carriers are electrons for the all $V_{\rm tg}$.
For the calculation of the velocity, the contribution of the ungated region to the time of flight is subtracted.
{\bf g}, The velocity around $\nu =6$ for $V_{\rm tg}=0$, 10, 20, and 30\ V.}
\label{gate}
\end{center}
\end{figure*}

\end{document}